\documentclass[10pt,times,a4paper]{iopart}
\expandafter\let\csname equation*\endcsname\relax
\expandafter\let\csname endequation*\endcsname\relax

\usepackage{times}
\usepackage[usenames]{color}
\usepackage{amsfonts}
\usepackage{amsmath}
\usepackage{amssymb}
\usepackage{bm}
\usepackage{dcolumn}
\usepackage{enumerate}
\usepackage{epsfig}
\usepackage{graphicx}
\usepackage{graphics}
\usepackage{multirow}
\usepackage[latin1]{inputenc}
\usepackage{latexsym}
\usepackage{rotating}
\usepackage{hyperref}
\usepackage[caption=false]{subfig}
\usepackage{float}
\usepackage[dvipsnames]{xcolor}
\usepackage{cite}

\begin{document}

\title[Ten years of tests of GR with GW observations]{Ten years of extreme gravity tests of general theory of relativity with gravitational-wave observations}

\author{Anuradha Gupta}
\address{Department of Physics and Astronomy, University of Mississippi, University, Mississippi 38677, USA}
\ead{agupta1@olemiss.edu}

\date{\today}

\begin{abstract}
Ten years ago, the first direct detection of gravitational waves (GWs) from the merger of two black holes, GW150914, provided the very first opportunity to test Einstein's general theory of relativity (GR) in the extreme gravity regime, where the gravitational field is strong, characteristic speeds are highly relativistic, and spacetime is dynamical. Such a regime is currently accessible only through coalescing compact binaries. In this review, we summarize the status of testing GR with GW observations and discuss the lessons learned. We also touch upon the challenges we currently have in testing GR and the potential path forward to detect a credible violation of GR, should one exist in the data.

\end{abstract}

\section{Introduction}

A century ago, Albert Einstein formulated the general theory of relativity (GR)~\cite{Einstein:1915ca}, redefining gravity as the warping of {\it spacetime} rather than a force. In this theory, gravity is governed by the distribution of mass-energy through Einstein's field equations, whose solutions describe various physical phenomena such as planetary motion, stellar evolution of stars, black holes, and large-scale dynamics of the universe. Since its conception, GR has been experimentally verified on numerous occasions --- for example, by correctly predicting the perihelion precession of Mercury~\cite{Einstein:1915bz}, the bending of light in curved spacetime~\cite{Dyson:1920cwa}, the gravitational redshift~\cite{Pound:1960zz}, the frame-dragging effects measured by Gravity Probe B~\cite{Everitt:2011hp}, and the emission of gravitational radiation from the Hulse-Taylor binary pulsar~\cite{Taylor:1979zz}. More recent tests through solar-system~\cite{Will:2014kxa}, binary pulsar~\cite{Kramer:2021jcw,Freire:2024adf}, and cosmological~\cite{Psaltis:2008bb} observations have also been successfully passed by GR.

Gravitational waves (GWs) are one of the key predictions of GR~\cite{Einstein:1918btx}. They are time-varying perturbations of spacetime that propagate away from their source at the speed of light. These waves are transverse in nature and possess two independent polarization states: plus ($h_+$) and cross ($h_\times$) polarizations. All the tests mentioned above probe regimes of gravity that are entirely different from those accessible through GWs from coalescing compact binaries. For instance, solar-system tests probe the {\it weak field} regime, where gravity is weak, the characteristic speeds of sources are small compared to the speed of light, and spacetime is effectively non-dynamical. Binary pulsar tests probe {\it strong field} regimes, where gravity is strong (since pulsars are highly self-gravitating), but the characteristic speeds remain small and spacetime is still approximately non-dynamical. It is coalescing compact binaries that provide access to the {\it extreme gravity} regime, where gravity is strong, characteristic speeds are relativistic, and spacetime is highly dynamical. Sampling this regime is only possible through direct GW observations. It is worth noting that direct imaging of supermassive black holes by the Event Horizon Telescope (EHT)~\cite{EventHorizonTelescope:2019dse,EventHorizonTelescope:2022wkp} can also probe strong gravitational-potential regimes~\cite{EventHorizonTelescope:2022xqj}. However, the sources observed by the EHT are quasi-stationary, and the spacetime curvature around supermassive black holes is much smaller than that around the stellar-mass black holes and neutron stars observed with ground-based GW detectors~\cite{LIGOScientific:2018mvr,LIGOScientific:2020ibl,LIGOScientific:2021djp,LIGOScientific:2025slb}. 

In 2015, the two LIGO detectors made the first direct detection of GWs from the coalescence of a binary black hole (BBH), GW150914~\cite{LIGOScientific:2016aoc}. This observation provided the first opportunity to test GR in the extreme gravity regime and to constrain its predictions~\cite{LIGOScientific:2016lio}, as well as those of alternative theories of gravity~\cite{Yunes:2016jcc}. Over the past decade, a total of 218 confident GW events have been detected by the LIGO-Virgo detector network~\cite{LIGOScientific:2025slb}. These include several exceptional events~\cite{LIGOScientific:2017ycc,LIGOScientific:2017vwq,LIGOScientific:2020stg,LIGOScientific:2020zkf,LIGOScientific:2020iuh,LIGOScientific:2024elc,LIGOScientific:2025cmm,LIGOScientific:2025rid,LIGOScientific:2025rsn,LIGOScientific:2025brd} whose unique properties have enabled particularly stringent tests of GR. In this paper, we review the first ten years of tests of extreme gravity using GW observations. The work is largely based on an invited talk presented at the American Physical Society's Global Physics Summit~\cite{APStalk}. We primarily focus on tests of GR performed by the LIGO-Virgo-KAGRA (LVK) Collaboration using real GW data from the LIGO~\cite{LIGOScientific:2014pky} and Virgo~\cite{VIRGO:2014yos} detectors. Where relevant, we also discuss results from the broader literature that have employed real GW data. This paper does not cover constraints on specific beyond-GR theories. For comprehensive reviews of tests of alternative gravity theories using various detectors and future prospects, we refer the reader to the following resources~\cite{Yunes:2013dva,Berti:2015itd,Berti:2018cxi, Berti:2018vdi,Berti:2019xgr,Barausse:2020rsu,LISA:2022kgy,Yunes:2024lzm,Berti:2024orb}.

The structure of this paper is as follows. Section~\ref{how_to_tests_of_gr} outlines the various approaches used to test GR with GW observations. Section~\ref{sec:data_analysis} summarizes the standard practices employed in data analysis and the reporting of results. In section~\ref{tests_of_gr}, we provide a more detailed discussion of the GR tests and present results primarily from BBH events. Section~\ref{sec:bns} focuses on tests (and corresponding results) involving binaries that contain neutron stars. Finally, section~\ref{sec:concl} concludes the paper and discusses future prospects for testing GR.

Note: The KAGRA detector~\cite{KAGRA:2020tym} joined the LIGO-Virgo network during the second half of the third observing run (O3b). Therefore, when referring to results prior to O3b, we use ``LIGO-Virgo Collaboration'' instead of ``LVK Collaboration."

\section{How to test a theory of gravity such as GR?}
\label{how_to_tests_of_gr}
There are two main approaches to testing a theory of gravity (such as GR) using GW observations: direct tests and generic tests. In a direct test, one compares the predictions of a beyond-GR theory, through an appropriate waveform model, with GW data. At present, not many mature beyond-GR theories with well-posed initial-value formulation exist, although some progress has been made for certain subclasses of scalar-tensor theories (see, e.g.,~\cite{Ohashi:1996uz, Brunetti:1998cc, Healy:2011ef, Barausse:2012da, Lang:2013fna, Palenzuela:2013hsa, Mirshekari:2013vb, Berti:2013gfa, Lang:2014osa, Sennett:2016klh, Sennett:2016rwa, Zhang:2017srh, Okounkova:2017yby, Bernard:2018hta,  Bernard:2018ivi, Bernard:2019yfz, Okounkova:2019dfo, Okounkova:2019zjf, Zhao:2019suc,  Figueras:2021abd, Brax:2021qqo, Guo:2021leu, Julie:2022qux, Higashino:2022izi, Bernard:2022noq, Bernard:2023eul, Kuan:2023hrh, Heisenberg:2023prj, Ma:2023sok, Lam:2024wpq,  Trestini:2024zpi, Almeida:2024uph}). However, these theories are still not sufficiently developed (as compared to GR) to be directly compared with GW data. On the other hand, in a generic test, one focuses on a specific feature of GR and examines whether that feature shows any deviation from the GR prediction when compared with GW data. This approach is often referred to as a {\it null test} of GR: Einstein's theory is the null hypothesis and possible deviations are searched for. A potential drawback of this approach is that, if a deviation is found, it cannot be uniquely mapped to a specific beyond-GR theory; rather, a class of alternative theories may be consistent with the observed deviation. In this paper, we only focus on null tests of GR.

Null tests can be classified according to which aspect of GW emission or propagation they probe. We classify them into the following categories:

\begin{itemize}
\item {\bf Consistency tests:} These tests search for possible deviation from GR by checking the consistency between the observed GW signal and the GR waveform (or other prediction). This method does not require any other assumption or introduction of phenomenological deviations to the GR waveform model. 

\item {\bf Generation tests:} The relationship between source properties and outgoing radiation can differ across alternative theories of gravity. For example, the presence of additional fields or higher-curvature corrections can change the binary's binding energy and angular momentum, as well as the corresponding fluxes of energy and angular momentum~\cite{Blanchet:2002av,Lang:2013fna,Lang:2014osa,Bernard:2018hta,Sotiriou:2006pq,Shiralilou:2021mfl,Julie:2018lfp,Yagi:2011xp,Mirshekari:2013vb}, resulting in a GW signal that deviates from GR predictions. Studying GW generation from the source requires solving the linearized field equations around a fixed background metric. This can be done either using a multipolar post-Newtonian (PN) formalism (see~\cite{Blanchet:2002av} for a review) or through fully numerical methods. However, this is a challenging problem and has been investigated only for a limited number of beyond-GR theories. In practice, therefore, parameterized modifications to GR waveforms are introduced, and these modifications are then tested for consistency with GR.

\item {\bf Propagation tests:} In GR, GWs propagate through the intervening spacetime between the source and Earth non-dispersively and without attenuation. In beyond-GR theories, however, GW propagation can exhibit effects such as dispersion~\cite{Mirshekari:2011yq}, birefringence~\cite{Okounkova:2021xjv}, and amplitude damping~\cite{Nishizawa:2017nef,Belgacem:2017ihm}. Typically, the tests performed within the LVK Collaboration assume that GW generation near the source is the same as in GR, and that any differences in the signal received on Earth arise solely from propagation effects. Such tests are particularly suited for probing theories such as massive graviton theories, where generation effects are suppressed by powers of $\lambda / \lambda_{g} \ll 1$, with $\lambda_g$ being the Compton wavelength of the graviton and $\lambda$ the gravitational wavelength. Moreover, even if generation effects are present in a beyond-GR theory, they are expected to be subdominant compared to propagation effects.

\item {\bf Polarization tests:} A generic metric theory of gravity can support up to six polarization modes: two tensor, two vector, and two scalar modes~\cite{Eardley:1973br,PhysRevD.8.3308}. In contrast, GR allows only two tensor modes: the plus and cross polarizations. Therefore, the detection of any vector or scalar polarization would indicate a deviation from GR.

\item {\bf Kerr nature tests:} GWs observed from compact-object binaries could also originate from systems involving compact objects other than the black holes predicted by classical GR. Several alternative theories of gravity predict the existence of {\it exotic} compact objects whose GW signals can closely resemble those from BBHs in GR. These exotic compact objects include boson stars~\cite{PhysRev.172.1331,PhysRev.187.1767}, fuzzballs~\cite{Mathur:2009hf,Bena:2022rna}, gravastars~\cite{Mazur:2004fk}, ultracompact anisotropic stars~\cite{Bowers:1974tgi,Raposo:2018rjn}, elastic stars~\cite{Alho:2022bki,Alho:2023mfc,Alho:2023ris}, wormholes~\cite{1988AmJPh..56..395M,Visser:1995cc,Lemos:2003jb}, strange quark stars~\cite{Haensel:1986qb,Alcock:1986hz}, asymmetric dark-matter stars~\cite{Kouvaris:2015rea}, and frozen stars~\cite{Brustein:2016msz}. The merger remnant of such binaries could be either a black hole or another exotic compact object. Hence, testing the nature of the components in the binary and of the merger remnant can constrain the gravity theories predicting them, or test the no-hair conjecture~\cite{Hansen:1974zz,Carter:1971zc,Gurlebeck:2015xpa}.
\end{itemize}

We note that the tests considered above are not fully independent from one another and may exhibit some overlap or redundancy. Studying the degeneracy between different tests and their outcomes is a complex and time-consuming process. Fortunately, some work~\cite{Johnson-McDaniel:2021yge} has started looking into this, but understanding the relation between various null tests of GR is vital for the sustainability of testing GR enterprise. However, it is important to note that these tests are also complementary to each other. For instance, theories that depart from GR in terms of generation will also depart in propagation. Similarly, theories that predict an exotic compact object as the merger remnant may also have inspiral dynamics different from GR. Hence, different types of deviations from GR may manifest as GR violations in multiple tests.

Further, these tests are not perfect and suffer from limitations in probing the underlying physics they are designed to test. First, the performance of all such tests depends critically on the accuracy of the waveform models employed, as discussed in section 3 of~\cite{Gupta:2024gun}. Current waveform models do not incorporate all known physics within GR and are also affected by numerical truncation errors. These shortcomings can lead to false identification of apparent GR violations in many tests that rely on waveform modeling. In addition, some of these tests are not yet capable of capturing the full range of possible beyond-GR effects. For instance, in their current implementation, generation tests probe deviations only up to 3.5PN order in the inspiral phase and may therefore miss modifications appearing at higher PN orders, such as the 5PN corrections predicted by higher-curvature gravity theories~\cite{Bernard:2025dyh}. Propagation tests, on the other hand, are more sensitive to sources at large distances; consequently, the limited horizon distance of current detectors restricts their constraining power. Furthermore, fully distinguishing between the six independent polarization modes requires a network of at least five detectors. With only three detectors currently taking GW data, the scope of polarization tests remains limited. Similarly, tests of the Kerr nature of the merger remnant mostly rely on the post-merger signal, which typically has low signal-to-noise ratios (SNRs), thereby reducing the sensitivity of such tests. Overall, these limitations can be mitigated with more accurate and complete waveform models, as well as with a larger network of more sensitive GW detectors.

\section{Note on Data Analysis Methods, Results Reporting Practices, and Selection Criteria}
\label{sec:data_analysis}
Almost all of the GR tests performed within the LVK Collaboration and the broader GW community are based on the Bayesian inference framework, in which we infer the parameters of a model that best describe the astrophysical signal in the data. Here, the model could either be GR or an alternative theory of gravity, but often it is a baseline GR model with additional non-GR parameters. In the Bayesian framework, we estimate the {\it posterior} probability --- the probability that the model with source parameters is responsible for the data. The posterior probability is defined as the product of {\it likelihood} of data given the source parameters and {\it prior} probability of model parameters. A model with multiple parameters gives rise to a multidimensional posterior, and one-dimension posterior for a given parameter (e.g., binary masses or a GR deviation parameter) is computed by marginalizing that multidimensional posterior over the rest of the model parameters. In practice, the posterior is stochastically sampled over the parameter space by using techniques like Metropolis-Hastings Markov-Chain Monte Carlo~\cite{Metropolis:1953am,Hastings:1970aa}, nested sampling~\cite{Skilling2004a}, or dynesty sampling~\cite{Speagle:2019ivv} while using the Bayesian inference packages such as \textsc{LALInference}~\cite{Veitch:2014wba} and \textsc{Bilby}~\cite{Ashton:2018jfp}.

The growing number of GW events calls for population-level statements about tests of GR, which require combining information from all analyzed signals. Moreover, small deviations that are not significant in individual events may become more apparent when considered collectively. Two main approaches are commonly used to combine results across multiple observations: (i) simple multiplication of likelihoods and (ii) hierarchical inference. The first method assumes that the deviation from GR is the same across all events in the population~\cite{Zimmerman:2019wzo}, allowing the likelihoods from individual events to be multiplied to obtain the posterior on the population-level deviation parameter. This assumption is rather strong, as deviations from GR may, in general, depend on the binary parameters and thus vary across events. In the second approach, the non-GR parameters for each event are assumed to follow a common underlying distribution (the {\it hyperdistribution}), which is determined by the true theory of gravity and the source population, after accounting for selection biases. In practice, a simple, minimally informative Gaussian distribution with unknown mean and standard deviation is used to model the hyperparameters~\cite{Isi:2019asy}. If the population is consistent with GR, the mean should agree with the GR prediction (i.e., for a null test of GR), and the standard deviation should be consistent with zero. In this paper, when quoting results from a population of sources, we report only those obtained using hierarchical Bayesian inference, except for propagation tests where simple multiplication of likelihoods is used. Unless otherwise specified, all quoted values correspond to the 90\% credible interval of the posterior distribution.

Due to the high computational cost associated with GR tests, only a subset of highly significant events with a false alarm rate (FAR) of less than $10^{-3}$ year$^{-1}$ are used to test GR. Additionally, only events detected by two or more interferometers qualify for tests of GR analyses. These selection criteria were not followed when analyzing the GWTC-1 events, which included only ten events at that time (see Table 1 of~\cite{LIGOScientific:2019fpa}). The GWTC-2 testing GR paper~\cite{LIGOScientific:2020tif} analyzed 24 additional events and GWTC-3 analyzed 15 additional events~\cite{LIGOScientific:2021sio} that satisfied this FAR threshold. Furthermore, different GR tests have their own selection criteria for events to ensure that the physics being probed is meaningful. Table~\ref{summary} summarizes the selection criteria for various tests performed in the literature.

\begin{table}[ht!]
\label{summary}
\centering
\resizebox{\columnwidth}{!}{%
\begin{tabular}{|c|p{5.5cm}|p{5.5cm}|}
\hline
\textbf{GR test} & \textbf{Selection Criteria} & \textbf{Remarks} \\ \hline
Residual Test & N/A &  \\ \hline
IMR Consistency Test & $\rho_{\rm insp}\geq6$, $\rho_{\rm postinsp}\geq6$, and $M_z<100M_\odot$ & The boundary between inspiral (insp) and post-inspiral (postinsp) is defined by the quadrupolar mode GW frequency of the innermost stable circular orbit of the remnant Kerr black hole; $M_z$ is the redshifted total mass of the binary. \\ \hline
meta-IMR Consistency Test & N/A &  \\ \hline
Eccentricity Evolution Consistency Test & Eccentricity posterior at the reference frequency excludes zero at 68\% credible interval  &  \\ \hline
TIGER & $\rho_{\rm insp}\geq6$ or $\rho_{\rm postinsp}\geq6$ & The boundary between inspiral and post-inspiral is defined by the minimum energy circular orbit frequency. \\ \hline
FTI & $\rho_{\rm insp}\geq6$ & Inspiral is defined up to $0.35$ times the frequency of the quadrupolar mode at the peak amplitude defined in SEOBNRv4. \\ \hline
PCA & $\rho_{\rm insp}\geq15$ & $\rho_{\rm insp}$ computed according to TIGER and FTI frameworks. \\ \hline
Parameterized post-Einsteinian Test & N/A &  \\ \hline
Modified Dispersion Relation Test & N/A & {\small GW191109\_010717, GW200316\_215756}, and {\small GW200115\_042309} were excluded from the GWTC-3 analysis due to issues with non-stationary noise, posterior sampling, and high computational cost, respectively.  \\ \hline
Polarization Test & N/A & {\small GW190425\_081805, GW190720\_000836, GW190828\_065509, GW191129\_134029, GW200115\_042309, GW200202\_154313}, and {\small GW200316\_215756} were excluded from the GWTC-3 analysis for not having a strong enough time-frequency track. \\ \hline
Spin Induced Quadrupole Moment Test & $\rho_{\rm insp}\geq6$, $\chi_{\rm eff}$ posterior excludes zero within 68\% credible interval & Inspiral regime is defined up to frequency $0.018 (c^3/GM_z)$. \\ \hline
pyRing Test & BBH remnant mass and spin are constrained relative to the prior and evidence favors a signal over Gaussian noise &  \\ \hline
pSEOBNRv4HM Test & $\rho_{\rm insp}\geq8$ and $\rho_{\rm postinsp}\geq8$ &  \\ \hline
Echoes Search & N/A &  \\ \hline
\end{tabular}%
}
\caption{Summary of tests of GR conducted within the LVK Collaboration and in the broader literature, and their selection criteria. In the GWTC-1 tests of GR paper~\cite{LIGOScientific:2019fpa}, a two-tier event selection criterion was adopted based on search significance: ${\rm FAR} < 1\,{\rm year}^{-1}$ for single-event analyses and ${\rm FAR} < 10^{-3}\,{\rm year}^{-1}$ for population-level analyses. However, due to the substantial computational cost associated with the increasing number of detections, only events with ${\rm FAR} < 10^{-3}\,{\rm year}^{-1}$ have been considered for GR tests from GWTC-2 onward~\cite{LIGOScientific:2020tif}. In addition, only events detected by at least two interferometers qualify for the test of GR analyses. The selection criteria have evolved for many tests over the years; the latest criteria are summarized here. The symbol $\rho$ denotes the detector network SNR.}
\label{tab:gr_tests}
\end{table}

\section{Tests of GR and results from GW observations}
\label{tests_of_gr}

\begin{figure*}[htb]
\centering
\includegraphics[width=\textwidth]{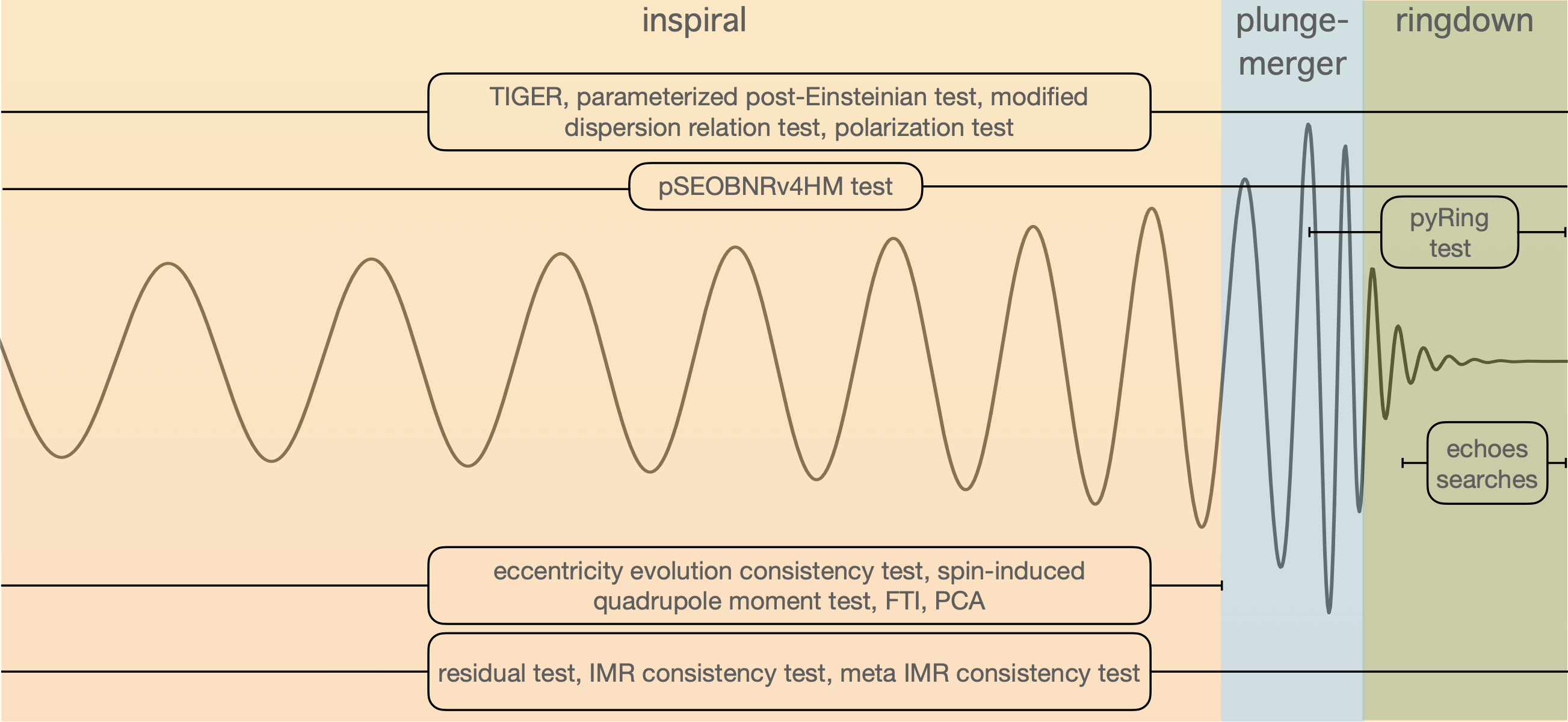}
\caption{\label{fig:tgr_summary}Illustration of the range of applicability of various tests of GR discussed in this paper. The figure shows a schematic GW signal from a typical BBH merger in the LIGO/Virgo sensitivity band, with a rough demarcation between the inspiral, plunge-merger, and ringdown phases, indicating where different tests are applicable. This figure is reproduced from Fig.~6 of~\cite{LIGOScientific:2025cmm}.}
\end{figure*}

In this section, we discuss each of the classes of GR tests mentioned in section~\ref{how_to_tests_of_gr} in more detail, list the specific tests that have been performed, and summarize their outcomes when applied to real GW signals. Different tests within a given class probe different parts of the GW signal. Figure~\ref{fig:tgr_summary} provides a schematic overview of the signal regions to which each test is applied.

\subsection{Consistency Tests}
Four types of consistency tests have been performed in the literature. The first of these is the {\it residual test}~\cite{LIGOScientific:2016lio,LIGOScientific:2019fpa,LIGOScientific:2020tif,LIGOScientific:2021sio}, which checks the overall consistency of the GW signal with the data. In this test, the best-fit GR waveform (computed at the maximum-likelihood parameters) is subtracted from each detector's data to determine whether the residuals exhibit any statistically significant excess power. The idea is that if the data are consistent with the GR signal, the residuals should be consistent with instrumental noise, which is assumed to be stationary and Gaussian. This test was first performed on GW150914~\cite{LIGOScientific:2016lio} by subtracting a signal reconstructed from a burst analysis~\cite{LIGOScientific:2016fbo} that does not rely on any theoretical model. The residuals were analyzed using the BayesWave~\cite{Cornish:2014kda} algorithm and were found to be statistically indistinguishable from the instrumental noise around GW150914. The test was later applied to events in GWTC-1~\cite{LIGOScientific:2019fpa}, GWTC-2~\cite{LIGOScientific:2020tif}, and GWTC-3~\cite{LIGOScientific:2021sio}, using waveform models \{IMRPhenomPv2~\cite{Khan:2015jqa,Husa:2015iqa,Hannam:2013oca}\}, \{IMRPhenomPv2, IMRPhenomPv3HM~\cite{Khan:2018fmp,Khan:2019kot}, NRSur7dq4~\cite{Varma:2019csw}\}, and \{IMRPhenomXPHM~\cite{Pratten:2020ceb,Pratten:2020fqn,Garcia-Quiros:2020qpx}\}, respectively, for subtraction. No statistically significant evidence of excess power in the residual data was observed for any event analyzed in GWTC-3.

The second of these is the {\it inspiral-merger-ringdown
(IMR) consistency test}~\cite{Ghosh:2016qgn,Ghosh:2017gfp}, which looks for the self-consistency of the signal. In this test, one checks for consistency between the low- and high-frequency portions of the signal by comparing the inferred final mass and spin of the merger remnant obtained by analyzing the low- and high-frequency portions of the signal. The final mass and spin of the remnant are computed using numerical relativity fits~\cite{Healy:2016lce,Hofmann:2016yih,Jimenez-Forteza:2016oae,spinfit-T1600168}. See~\cite{Ghosh:2017gfp} for a detailed description of the IMR consistency test and how the boundary of low- and high-frequency portions is selected. Similar to the residual test, this test was also first applied to GW150914.  The low- and high-frequency estimates for final mass and spin were found to be consistent with each other. 

Fast forward, IMR consistency test has been routinely applied to events in GWTC-1, GWTC-2, and GWTC-3, while using progressively accurate waveform models:   IMRPhenomPv2 for GW150914 and GWTC-1, IMRPhenomPv2 and IMRPhenomPv3HM for GWTC-2, and IMRPhenomXPHM for GWTC-3. All events in GWTC-2 passed this test except for GW170823 and GW190814. These two binaries had their GR quantiles\footnote{defined as the fraction of the reweighted posterior probability contained within the isoprobability contour intersecting the GR value.} among the largest (i.e., very small consistency with GR), which could be attributed to their relatively low SNRs in each portion of the signal (both low- and high-frequency portions for GW170823 and only high-frequency portion for GW190814). The additional events added to the GWTC-3 analysis (with additional constraint, see Table~\ref{tab:gr_tests}) were all consistent with GR. The most recent results of this test are reported in~\cite{LIGOScientific:2025obp}, where a suite of GR tests is applied to GW250114~\cite{LIGOScientific:2025rid}, the loudest BBH merger observed to date with network SNR $\sim 80$. Remarkably, the constraints on deviation parameters inferred from GW250114 alone are comparable to those obtained from the combined analysis of $30$ events in GWTC-4.

\begin{figure*}[htb]
\centering
\includegraphics[width=0.8\textwidth,trim = {0 30 0 30}]{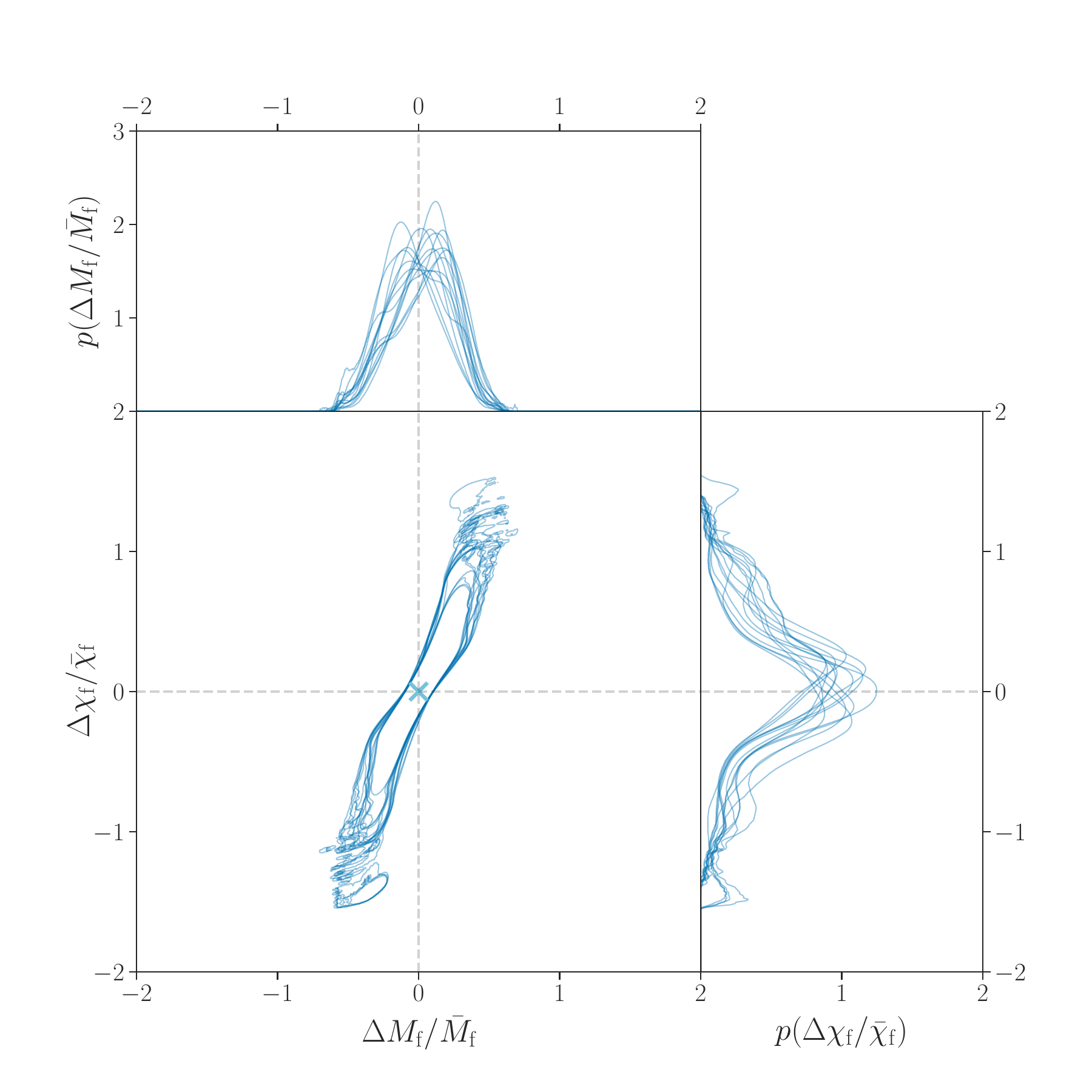}
\caption{\label{fig:meta_imrct}Results of the meta-IMR consistency test for GW190521. Shown are the two-dimensional posterior probability distributions of the deviation parameters $\Delta M_{\rm f}/M_{\rm f}$ and $\Delta \chi_{\rm f}/\chi_{\rm f}$, along with the corresponding one-dimensional marginalized posteriors. For this event, only the usual parameter-estimation analysis and data from only one type of parameterized test were available, yielding a total of 15 analysis pairs for the meta-IMR consistency test. The maximum GR quantile across all pairs was $11.4$, suggesting strong consistency between GW190521 data and GR. For more details, see section 5.4 of~\cite{Madekar:2024zdj}. This figure is regenerated from data produced in~\cite{Madekar:2024zdj}.}
\end{figure*}

Johnson-McDaniel et al.~\cite{Johnson-McDaniel:2021yge} found that the final mass and spin posteriors obtained from different parameterized tests, or from a standard parameter estimation analysis, can differ substantially when applied to a non-GR signal. This shows that different GR tests can bias the inferred final mass and spin in different ways, even when applied to the same signal. This observation led to the development of a new test, {\it the meta-inspiral-merger-ringdown (meta-IMR) consistency test}~\cite{Madekar:2024zdj}. This test is based on the original IMR consistency test above, but instead of checking the consistency between the low- and high-frequency portions of the signal, the meta IMR consistency test checks for consistency between different parameterized tests of GR (or the usual parameter estimation analysis) performed on the same signal (see section~3 of~\cite{Madekar:2024zdj} for more detail). For example, if one performs parameterized tests~\cite{Meidam:2017dgf, Agathos:2013upa, Roy:2025gzv,Mehta:2022pcn} (discussed in section~\ref{sec:generation_test}) on a signal by varying one deformation parameter at a time, meta-IMR consistency test will check for the consistency between the final mass and spin inferences in pairs of parameterized tests with different deformation parameters. Madekar et al.~\cite{Madekar:2024zdj} applied the meta-IMR consistency to a few selected events from GWTC-3 (GW170817~\cite{LIGOScientific:2017vwq}, GW190412~\cite{LIGOScientific:2020stg}, GW190521~\cite{LIGOScientific:2020iuh}, GW190814~\cite{LIGOScientific:2020zkf}, GW200225\_060421~\cite{LIGOScientific:2021djp}), and these all were found to be consistent with GR. As an example, in figure~\ref{fig:meta_imrct} we show the results of the meta-IMR consistency test applied to the available analyses of GW190521.

Recently, Bhat et al.~\cite{Bhat:2025lri} introduced an {\it eccentricity evolution consistency test} designed to identify eccentric signatures in BBH signals. The test can also indicate whether apparent eccentricity is instead being mimicked by other GR or beyond-GR effects in quasi-circular binaries. This test is similar to the IMR consistency test: one first infers the eccentricity and other binary parameters at a (lower) reference frequency, and then predicts the eccentricity at higher frequencies assuming GR. If the signal exhibits the eccentricity evolution as expected in GR, the eccentricity inferred at multiple higher frequencies should be consistent with the GR prediction. This test was applied to GW200105~\cite{LIGOScientific:2021qlt}, a neutron-star -- black-hole binary candidate, which passed the consistency check, indicating that the signal is compatible with an eccentric binary as described by GR~\cite{Tiwari:2025fua}.

\subsection{Generation Tests}
\label{sec:generation_test}
In order to test whether the generation of GWs from the source is consistent with the GR prediction, two main approaches are used in the literature: (i) {\it the parameterized test of post-Newtonian theory} and (ii) {\it the parameterized post-Einsteinian test}.   

\paragraph{Parameterized test of post-Newtonian theory:} The idea of the parameterized test of post-Newtonian theory (hereafter parameterized test) was first proposed by Arun et al.~\cite{Arun:2006yw}, which was later developed and refined by several authors~\cite{Mishra:2010tp,Meidam:2017dgf,Agathos:2013upa,Roy:2025gzv,Mehta:2022pcn}. In this framework, fractional deviation parameters $\delta \hat{p}_i$ are inserted into the phase of a baseline GR waveform model such that $p_i\rightarrow p_i(1+\delta \hat{p}_i)$, where $p_i$ represents the PN coefficients as well as the phenomenological coefficients in the late-inspiral and merger phases of the waveform (see Table I in~\cite{LIGOScientific:2016lio} for a summary of these fractional deviation parameters).  These parameters also include deviations at $-1$PN and $0.5$PN order, where there are no GR contributions, and are normalized by 0PN coefficients. Further, the $\delta \hat{p}_i$ are only added to the non-spinning terms of the PN phase to avoid catastrophic cancellation between phase terms due to the presence of spins. If the data are consistent with GR, the posterior on all $\delta \hat{p}_i$ should be consistent with zero. The LVK analysis usually employs two pipelines to perform the parameterized test: the Test Infrastructure for General Relativity (TIGER)~\cite{Meidam:2017dgf,Agathos:2013upa,Roy:2025gzv} and the Flexible-Theory-Independent (FTI)~\cite{Mehta:2022pcn}. The differences between the two pipelines are the following. TIGER constrains deviations in inspiral, late-inspiral, and merger phases whereas FTI focuses only on the inspiral part by tapering the deviations to zero beyond a certain frequency~\footnote{which is $0.35$ times the GW frequency of the quadrupolar mode at the peak of the amplitude as defined in the SEOBNRv4~\cite{Bohe:2016gbl} waveform model.} instead of allowing them to affect the rest of the signal. The baseline GR waveform model for TIGER are phenomenological waveforms~\cite{Khan:2015jqa,Garcia-Quiros:2020qpx} whereas FTI currently uses effective-one-body waveform models~\cite{Bohe:2016gbl,Cotesta:2018fcv} as baseline, though, in principle, it can work with any waveform model. The latest version of TIGER uses a precessing model with higher harmonics (IMRPhenomXPHM), whereas FTI currently employs a non-precessing model with higher harmonics (SEOBNRv4HM\_ROM~\cite{Bohe:2016gbl,Cotesta:2018fcv,Cotesta:2020qhw}).

GW150914 provided the first constraints on higher-order PN deviation parameters.\footnote{Binary pulsars, e.g., PSR J0737-3039~\cite{Kramer:2021jcw}, also provide constraints on PN deviation parameters, but these are generally weak except for the $-1$PN, $0$PN, and $0.5$PN coefficients. $-1$PN and $0$PN coefficients are constrained several orders of magnitude better while the 0.5PN coefficient yields a comparable constraint~\cite{Kramer:2021jcw}. These remarkably tighter bounds at $-1$PN and $0$PN order arise from the long observational time of binary pulsars.} Only the TIGER framework was applied in this analysis, as FTI had not yet been developed at that time. In this analysis, both single-parameter and multi-parameter tests were performed. In single-parameter tests, only one deviation parameter is allowed to vary (along with the binary's GR parameters), while the remaining deviation parameters are fixed to their GR values (i.e., zero). In these tests, deviations at multiple PN orders can still be detected, even if the phase at a specific PN order is not directly modified (see~\cite{Meidam:2017dgf,Johnson-McDaniel:2021yge}). In a multi-parameter test, all deviation parameters (as a group of inspiral, intermediate, and merger deviation parameters) were varied simultaneously along with the binary's GR parameters. Due to the strong correlations between parameters, the multi-parameter test did not provide meaningful constraints on most of the deviation parameters. However, all deviation parameters were found to be consistent with zero in both types of tests. After GW150914, only single-parameter parameterized tests have been performed on subsequent detections. The TIGER framework was applied to GW151226~\cite{LIGOScientific:2016sjg}, GW170104~\cite{LIGOScientific:2017bnn}, and GW170817 (along with FTI for the latter). Again, no statistically significant deviations from the GR predictions were found in any of these analyses. In the future, it will be possible to constrain all PN deviation parameters simultaneously using multiband observations of a population of BBHs with space- and ground-based GW detectors~\cite{Gupta:2020lxa,Datta:2020vcj}. Furthermore, the use of principal component analysis (PCA)~\cite{Pai:2012mv,Shoom:2021mdj} can provide tight constraints on combinations of PN deviation parameters even with a single-band observation~\cite{Saleem:2021nsb,Datta:2022izc,Datta:2023muk}. In fact, very recently, Mahapatra et al.~\cite{Mahapatra:2025cwk} applied PCA-based tests (in both the TIGER and FTI frameworks) to the GWTC-3 events and found them to be consistent with GR. 

In GWTC-1, the parameterized tests were applied to ten events, which were divided into two groups. Inspiral deviation parameters were tested only for events with an SNR in the inspiral regime of $\geq6$. These events were GW150914, GW151226, GW170104, GW170608, and GW170814. The TIGER framework using IMRPhenomPv2 and the FTI framework using SEOBNRv4\_ROM were applied to these events. On the other hand, only the intermediate and merger-ringdown parameters were tested for events with an SNR of $\geq6$ in the post-inspiral regime. These events were GW150914, GW170104, GW170608, GW170809, GW170814, and GW170823, and only TIGER using IMRPhenomPv2 was applied. The results obtained from both tests and all events were found to be consistent with the GR prediction. The same procedure was followed to analyze the 32 events of GWTC-2 (see Table V in~\cite{LIGOScientific:2020tif}). Similar to GWTC-1, IMRPhenomPv2 for TIGER and SEOBNRv4\_ROM for FTI were used to analyze signals. However, for three exceptional events, GW190412, GW190521, and GW190814, where the effect of higher-harmonics could be important, waveforms with higher-harmonics, IMRPhenomPv3HM and SEOBNRv4HM\_ROM, were used for TIGER and FTI, respectively, to analyze those signals. Again, the results were consistent with GR, and no GR deviations were found at any PN order. We note that the binary neutron star events GW170817 and GW190425~\cite{LIGOScientific:2020aai} were not included when combining results in the GWTC-1 and GWTC-2 analyses. Only FTI was applied to events in GWTC-3 because of the delays due to the upgrade of TIGER with the IMRPhenomX model. Fifteen additional events, which satisfied the FTI selection criteria, were analyzed in GWTC-3. There was no uniform improvement in the combined bound across all PN deviation parameters, but the resulting improvement was consistent with the increase in the number of events in the combined analysis. Later, TIGER results on GWTC-3 were presented in Roy et al.~\cite{Roy:2025gzv}, which used IMRPhenomXPHM to analyze the signals. No deviations from GR were found in this work either. 

\begin{figure*}[htb]
\centering
\includegraphics[width=\textwidth]{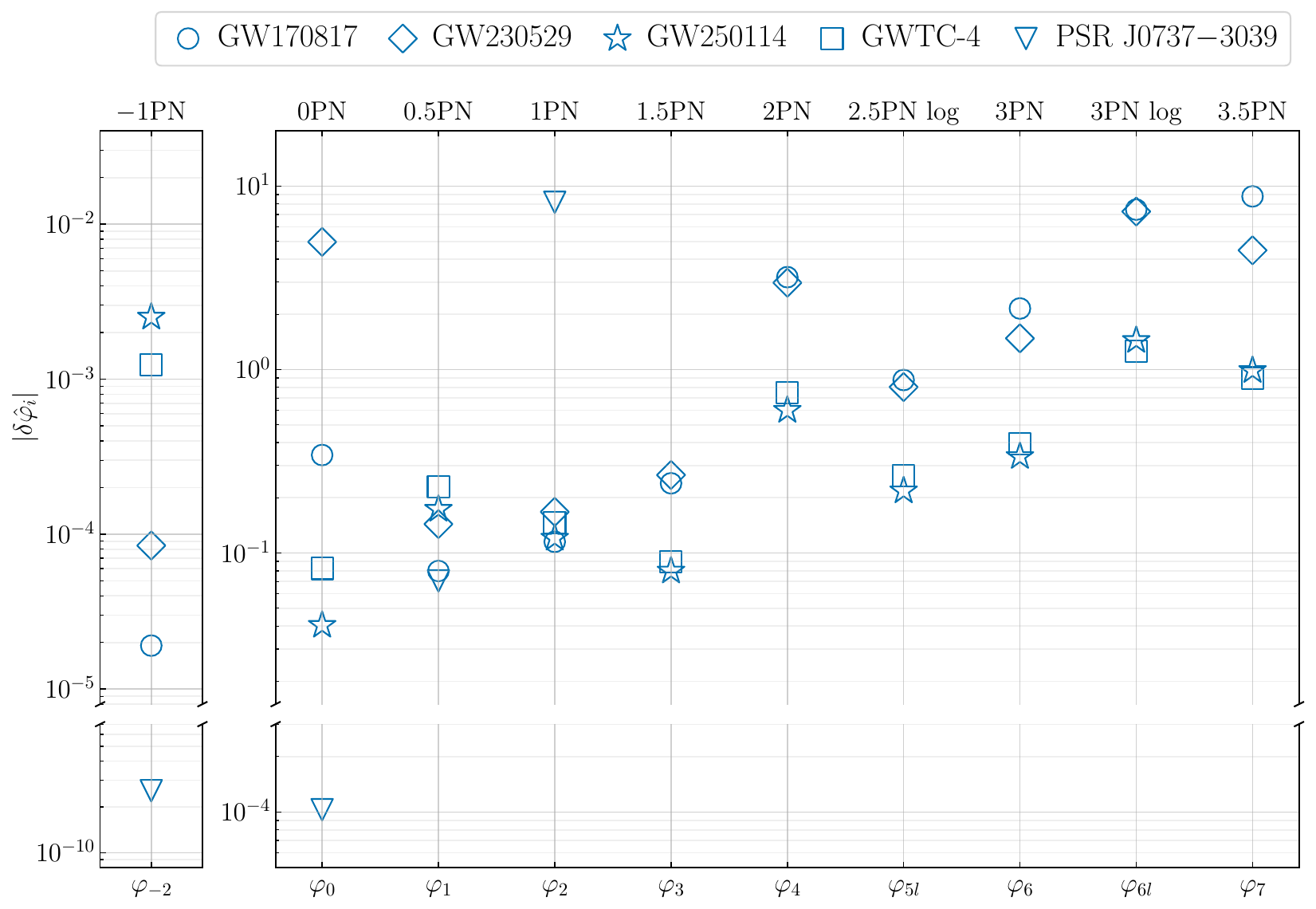}
\caption{\label{fig:tiger} Bounds on the fractional deviation parameters in the inspiral phase ($\delta \hat{\varphi}_i$) obtained from TIGER analyses of GW events and from the binary pulsar PSR J0737-3039~\cite{Kramer:2021jcw}. The horizontal axis denotes the PN coefficients (and corresponding PN orders), while the vertical axis shows the magnitude of the fractional deviation parameters, $|\delta \hat{\varphi}_i|$. In the legend, GW230529\_181500 has been shortened to GW230529 due to brevity. This figure is produced using public data and code from~\cite{LIGOScientific:2025obp,ligo_scientific_collaboration_2025_17018009}.}
\end{figure*}

In all the analyses discussed above, the tightest constraints were obtained for $-1$PN deviation parameter (which corresponds to the dipole radiation). The current bound from GWTC-3 is $7.3 \times 10^{-4}$ with the FTI analysis, which is about 19 times weaker than the bounds from the GW230529\_181500 alone~\cite{LIGOScientific:2024elc, Sanger:2024axs}, which itself is four times weaker than the bounds from GW170817~\cite{LIGOScientific:2018dkp} alone. However, none of these bounds are strong compared to those from binary pulsars, which constrain the $-1$PN deviation parameter to $\sim10^{-10}$~\cite{Kramer:2021jcw}. GW250114~\cite{LIGOScientific:2025rid} provided inspiral deviation parameter (i.e.,  $0$PN and higher order) bounds two to three times tighter than those from the full GWTC-4 catalog~\cite{LIGOScientific:2025obp}. In figure~\ref{fig:tiger}, we present bounds on inspiral (fractional) deviation parameters from $-1$PN to $3.5$PN for the exceptional events GW170817, GW230529\_181500~\cite{Sanger:2024axs}, and GW250114, and compare them with bounds from GWTC-4 and the binary pulsar PSR J0737-3039. For simplicity, only TIGER results are shown for GW events. The bounds on the $1.5$PN and higher PN order parameters from PSR J0737-3039 are not well constrained and are therefore not shown.

\paragraph{Parameterized post-Einsteinian test:} Yunes and Pretorius~\cite{Yunes:2009ke} argued that the deviations introduced in the parameterized test of post-Newtonian theory do not stem from any theoretical model of beyond-GR theory. Hence, to mitigate this issue, they proposed a more general formalism, parameterized post-Einsteinian test, that rewrites each harmonic of a GR waveform such that $h\rightarrow h_{\rm GR}(1+\alpha_{\rm ppE} u^{\rm a_{\rm ppE}})\,e^{i\beta_{\rm ppE}b_{\rm ppE}}$, where $u=(\pi {\mathcal M}f)^{1/3}$ is the reduced frequency parameter and ${\mathcal M}$ the chirp mass of the binary. Here $\alpha_{\rm ppE}$, $a_{\rm ppE}$, $\beta_{\rm ppE}$, and $b_{\rm ppE}$ are all free parameters that are varied along with the usual binary parameters in the data analysis. The possible values that these free parameters can take depend on the theory of gravity, e.g.,  $\alpha_{\rm ppE}=0\,, \beta_{\rm ppE}=0$ in GR. Table 4 in~\cite{Yunes:2024lzm} provides the values of $\alpha_{\rm ppE}$, $a_{\rm ppE}$, $\beta_{\rm ppE}$, and $b_{\rm ppE}$ from the known beyond-GR theories. It should be noted that the leading order PN corrections to the GR waveform can be mapped to the predictions of a few beyond-GR theories of gravity; however, it is argued that all such predictions from all known theories of gravity could be modeled in the parameterized post-Einsteinian formalism.

The first application of parameterized post-Einsteinian test on real GW data was on GW150914 and GW151226 in~\cite{Yunes:2016jcc}, where the authors extended the analyses in~\cite{LIGOScientific:2016lio,LIGOScientific:2016sjg}, and used the inferences to constrain several physical mechanisms taking place during the generation and propagation of GWs. For instance, they constrained the predictions of beyond-GR theories that include the activation of scalar fields, gravitational leakage into large extra dimensions, variable Newton's constant, modified dispersion relation, gravitational Lorentz violation, and the strong equivalence principle. Using GW150914 and GW151226, the authors also computed the bounds on the parameterized post-Einsteinian parameters while including negative PN corrections to the phase and reported that the bounds at negative PN and 1PN orders from GW observations are much larger than those obtained from binary pulsar or solar system observations, respectively (see Fig.~5 in~\cite{Yunes:2024lzm} for updated results). Later, Nair et al.\cite{Nair:2019iur} focused on the two low mass GWTC-1 events, GW151226
and GW170608~\cite{LIGOScientific:2017vox}, and used inferred values of PN deviation parameters from~\cite{LIGOScientific:2019fpa} to place stringent constraints on the predictions from higher-curvature theories, such as Einstein-dilaton-Gauss-Bonnet~\cite{Kanti:1995vq,Maeda:2009uy,Sotiriou:2013qea,Yagi:2015oca} and dynamical Chern-Simons gravity~\cite{Alexander:2009tp}, in the small-coupling approximation.

\subsection{Propagation Tests}
One of the predictions of massive graviton theories is that the generation of GWs is very close to that of GR but their propagation is different than that in GR. This is because if the graviton has mass, its propagation speed will differ from the speed of light, leading to dispersion of GWs. This can be generalized to Lorentz-violating gravity theories~\cite{Kostelecky:2016kfm,Calcagni:2009kc,Amelino-Camelia:2002cqb,Horava:2009uw,Sefiedgar:2010we}, where the dispersion relation can be parametrically modeled as~\cite{Mirshekari:2011yq}
\begin{equation}
    E^2= p^2c^2 + A_\alpha p^\alpha c^\alpha\,.
\end{equation}
Here $E$ and $p$ are the energy and momentum of the graviton, respectively. The phenomenological parameter $\alpha$ controls the frequency dependence and $A_\alpha$ controls the amplitude of dispersion. The values of $\alpha$ and $A_\alpha$ depend on the beyond-GR theory (see discussion in~\cite{Mirshekari:2011yq}). For example, $\alpha=0$ and $A_\alpha>0$ corresponds to the dispersion of a massive graviton with mass $m_g = \sqrt{A_0}/c^2$~\cite{Will:1997bb}. Similarly, $\alpha=2.5$ corresponds to  multi-fractal spacetime predictions~\cite{Calcagni:2009kc}, $\alpha=3$ to doubly special relativity~\cite{Amelino-Camelia:2002cqb}, $\alpha=4$ to Ho{\v{r}}ava-Lifshitz~\cite{Horava:2009uw}, extra-dimensional~\cite{Sefiedgar:2010we}, and standard model extension (with only non-birefringent terms)~\cite{Kostelecky:2016kfm} theories. In practice, the modified dispersion relation introduces a phase shift in the GW signal, whose magnitude depends on the parameters $\alpha$ and $A_\alpha$ along with the binary's mass and luminosity distance. To account for this effect in the signal observed on Earth, the corresponding dephasing is incorporated into a baseline GR waveform model.

The investigation of deviations from the standard dispersion relation ($E^2=p^2c^2$) within the LVK Collaboration has evolved quite a lot since the first analysis for GW150914~\cite{LIGOScientific:2016lio}. The GW150914 paper only considered a massive graviton theory with modified dispersion relation $E^2=p^2c^2+m_g^2c^4$ and tried to constrain the graviton's Compton wavelength $\lambda_g = h/(m_g c)$, where $h$ is the Planck constant. This Compton wavelength can also be translated into the characteristic length scale $\lambda_g$ for the Yukawa potential: $\phi(r) = (GM/r)[1 -\exp(-r/\lambda_g)]$, due to a body of mass  $M$ at a distance $r$. Note that in GR, $\lambda_g=\infty$ and $m_g=0$. Introducing dephasing due to dispersion at $1$PN order~\cite{Will:1997bb,Keppel:2010qu} of the SEOBNRv2\_ROM\_DoubleSpin~\cite{Purrer:2014fza,Purrer:2015tud} and IMRPhenomPv2 waveforms, the analysis of GW150914 constrained $\lambda_g > 1013 \, {\rm km}$, which corresponds to a graviton mass $m_g \leq 1.2 \times 10^{-22} \,{\rm eV/c^2}$. For the GWTC-1 analysis, IMRPhenomPv2 was used as the baseline GR model and each event's data were analyzed for $\alpha=\{0, 0.5, 1, 1.5, 2.5, 3, 3.5, 4\}$ while allowing for only positive and only negative values of $A_\alpha$. The case of $\alpha=2$ was excluded because the modification to the binary phase due to $\alpha=2$ is degenerate with an overall time delay of the signal in the detector. The results from positive and negative $A_\alpha$ were then combined to estimate the posterior on $A_\alpha$ for each $\alpha$ value (See Fig.~6 in~\cite{LIGOScientific:2019fpa}). After combining results from all GWTC-1 events, a combined bound on the graviton mass $m_g \leq 4.7 \times 10^{-23} \,{\rm eV/c}^2$ was reported. The same procedure was applied in GWTC-2 analysis, and a combined bound on $m_g \leq 3.09 \times 10^{-23} \, {\rm eV/c}^2$ was reported. The only change in the GWTC-3 analysis was that the IMRPhenomXP~\cite{Pratten:2020ceb} waveform model was used to analyze the data. The combined bound on $m_g \leq 2.42 \times 10^{-23} \,{\rm eV/c}^2$ was found, which is $1.31$ times better than the bounds of $m_g < 3.16 \times 10^{-23} {\rm eV/c}^2$ from Solar System observations~\cite{Bernus:2020szc}.

Niu et al.~\cite{Niu:2022yhr} used a modified dispersion relation [see Eq. (2.3) in~\cite{Niu:2022yhr}] that accounts for anisotropy and birefringence, along with the dispersion of GWs in the gauge-invariant linearized gravity sector of the Standard-Model Extension, and analyzed 50 events from GWTC-3. They did not find any evidence for Lorentz violation for the mass dimension $d = 5, 6$ cases (see also Haegel et al.~\cite{Haegel:2022ymk} for similar constraints on Lorentz- and Charge-Parity-Time-symmetry-violating effects). Later, Gong et al.~\cite{Gong:2023ffb} derived similar constraints on non-birefringent dispersions while analyzing all 90 events in GWTC-3 for the mass dimension $d = 6$ case. Zhu et al.~\cite{Zhu:2023rrx} analyzed 88 events from GWTC-3 and derived constraints on parity- and Lorentz-violating parameters while considering only isotropic GW propagation. Subsequently, Wang et al.~\cite{Wang:2025fhw} analyzed the same 88 events from GWTC-3, but this time constrained the effects of both Lorentz and diffeomorphism violations together on GW propagation in linearized gravity within the framework of the Standard-Model Extension for the mass dimension $d = 2, 3$ cases. They found no significant evidence for diffeomorphism violations in the GW data.

\subsection{Polarization Tests}
Multiple detectors with different orientations respond differently to a transient GW signal from a given sky location. This is because each detector is sensitive to a different linear combination of the polarization modes present in the signal. Hence, to fully distinguish among the six possible polarization modes, at least five GW detectors are required~\cite{Chatziioannou:2012rf}. Since GW150914 was detected by only two LIGO detectors (Hanford and Livingston), which have similar orientations, it was difficult to distinguish between GR and beyond-GR polarization modes. Despite this limitation, as an illustration, the Bayes factor was calculated between two hypotheses: one assuming the signal contains purely scalar modes and the other purely tensor modes. This type of hypothesis testing makes strong assumptions about the signal's polarization content and should be interpreted merely as a null test, as no viable theory of gravity currently predicts purely scalar or purely vector polarization modes in GWs. For GW150914, the waveforms and corresponding power spectral densities were reconstructed under purely tensor and purely scalar hypotheses using BayesWave. The resulting log Bayes factors were low, indicating no significant preference for either hypothesis.

GW170814~\cite{LIGOScientific:2017ycc} was the first event detected by three instruments (the two LIGOs and Virgo). Using the same hypothesis-testing method~\cite{Isi:2017fbj}, and employing IMRPhenomPv2 as the GR template, the analysis found that purely tensor polarization modes were strongly favored over purely scalar or vector modes. This result is consistent with the predictions of GR. Similar conclusions were drawn from GW170817~\cite{LIGOScientific:2018dkp}, which was also detected by three detectors. In this case, there was even stronger support for purely tensor polarization modes over purely scalar or vector modes. After GW170814, polarization tests were applied only to events detected by at least three detectors. In the GWTC-1 analysis, three additional events (GW170729, GW170809, and GW170818) satisfied this criterion, in addition to GW170814 and GW170817. Of these, only GW170818 had sufficient SNR and a small enough sky area to provide meaningful results. Once again, purely tensor polarization modes were favored over purely scalar or vector modes. We note that GW170817 provided the strongest evidence supporting pure tensor polarizations because of its precise sky localization enabled by the electromagnetic counterpart~\cite{LIGOScientific:2018dkp}. Later, Takeda et al.~\cite{Takeda:2020tjj} reanalyzed GW170814 and GW170817 using an improved framework for pure polarization modes, which accounted for nontensorial inclination dependence and incorporated the corresponding radiation patterns. For GW170814, their results were consistent with those reported by the LIGO-Virgo analysis~\cite{LIGOScientific:2017ycc}. In contrast, for GW170817, they found significantly stronger support for the purely tensorial hypothesis compared to the purely vector or purely scalar hypotheses, as compared to the LIGO-Virgo results~\cite{LIGOScientific:2018dkp}. This improvement arose from the use of a sky-location prior based on NGC 4993 and inclusion of binary-orientation information inferred from the gamma-ray burst jet (see also~\cite{Hagihara:2019ihn} for an independent analysis of GW170817).

For the GWTC-2 analysis, the same hypothesis testing was performed, but this time using a null stream technique~\cite{Guersel:1989th}, which does not require a specific waveform model. In this method, a linear combination of the data streams (the null stream) from different detectors is constructed for a given set of polarization modes and a specified sky location, such that the combination contains no signal and is consistent with noise. If, however, the true signal possesses polarization modes or a sky location different from those assumed, the null stream will not be consistent with noise and will indicate the presence of additional or different modes in the signal. The polarization test using null streams was applied to 20 GWTC-2 events, and none of the events favored either purely scalar or purely vector polarization modes over the GR tensor modes.

For the GWTC-3 analysis, an improved method~\cite{Wong:2021cmp} was used to search for evidence of mixed polarization modes in the data. Unlike the previous approach, which could test only two polarization modes at a time and required events detected by at least three detectors, this method can be applied to all events observed by two or more detectors. In this new technique, an effective antenna pattern function is constructed by selecting a subset $L$ of polarization modes and projecting the polarization state being tested into the corresponding subspace~\cite{Wong:2021cmp}. Each polarization mode can then be represented as a linear combination of the 
basis modes plus an additional orthogonal component. For all analyzed events in GWTC-3, both $L=1$ and $L=2$ were used. The choice $L=1$ is sufficient to capture the two tensor polarization modes since, in the quadrupolar approximation, the plus and cross modes differ only by a relative amplitude and phase that can be marginalized over when computing the Bayesian evidence. It was found that with $L=1$, the pure-scalar, pure-vector, and vector-scalar mixed hypotheses are strongly disfavored, while any mixed hypothesis containing tensor modes cannot be ruled out. Furthermore, the $L=2$ case revealed that mixed hypotheses can be more strongly disfavored than the pure-vector hypothesis, because the mixed models include a larger number of free parameters, resulting in a greater Occam penalty. Moreover, since the longitudinal and breathing modes are not linearly independent for interferometers, the pure-scalar hypothesis cannot be tested. Nevertheless, the GWTC-3 events were found to be consistent with the pure-tensor hypothesis.

\subsection{Kerr Nature Tests}
These tests are applied to check whether both the compact objects in a binary system or the merger remnant are consistent with a black hole or not. For the former, a test based on the measurement of the spin-induced quadrupole moments of the binary components is applied, while for the latter, methods based on black hole spectroscopy and echoes searches are employed. We will discuss these tests one-by-one below. 

\subsubsection{Inspiral test: Spin-induced quadrupole moment}
When a compact object is spinning, a series of spin-induced multipole moments is generated. The dominant mode is the spin-induced quadrupole moment which is defined as $Q = - \kappa M^3 \chi^2$, where $M$ is the mass, $\chi = S/M^2$ is the dimensionless spin parameter, and $S$ is the spin angular momentum of the compact object. The coefficient $\kappa$ is the dimensionless quadrupole parameter that quantifies the amount of distortion in the gravitational field outside the object due to its spin. For Kerr black holes $\kappa = 1$ \cite{Hansen:1974zz,RevModPhys.52.299}, whereas for slowly spinning neutron stars it is between $\sim$2 and $\sim$14 \cite{Pappas:2012qg,Pappas:2012ns,Harry:2018hke}. For spinning boson stars it is between $\sim$10 and $\sim$150 \cite{PhysRevD.55.6081} and for gravastars it could be negative \cite{Uchikata:2016qku}. The effect of spin-induced quadrupole moment of the compact objects is parameterized in terms of $\kappa_{1,2}$ in the waveform models derived from the PN approximation. For example, the leading order (mass-type quadrupole) spin-induced multipole moment first appears as a 2PN phase correction in the waveform and the PN corrections to this appear at 3PN and 3.5PN orders~\cite{Marsat:2014xea,Mishra:2016whh}. The sub-leading (current-type octupole) spin-induced multipole moment starts to contribute to the phase at the 3.5PN order~\cite{Marsat:2014xea}. 

In this test, deviation parameters $\delta \kappa_1$ and $\delta \kappa_2$ are introduced in the waveform phase such that $\kappa_1\rightarrow1+\delta \kappa_1$ and $\kappa_2\rightarrow1+\delta \kappa_2$~\cite{Krishnendu:2017shb,Krishnendu:2019tjp}. For a BBH, $\delta \kappa_1=0=\delta \kappa_2$, whereas $\delta \kappa_1\neq 0$ and/or $\delta \kappa_2 \neq0$ for a binary involving objects other than black holes. Hence, non-zero values of  $\delta \kappa_1$ and/or $\delta \kappa_2$ hint towards the presence of an exotic compact object in the binary. However, in practice, we measure the symmetric and antisymmetric combinations
$\delta \kappa_s = (\delta \kappa_1 + \delta \kappa_2)/2$ and
$\delta \kappa_a = (\delta \kappa_1 - \delta \kappa_2)/2$ instead. This is because of the correlation between mass and spin parameters which makes the measurement of $\delta \kappa_1$ and $\delta \kappa_2$ challenging. Again, for a BBH, both $\delta \kappa_s$ and $\delta \kappa_a$ vanish, whereas for a non-BBH system, one or both of these parameters can be non-zero. 

It was noted in~\cite{Krishnendu:2017shb} that for current generation GW detectors it is difficult to simultaneously measure $\delta \kappa_s$ and $\delta \kappa_a$. Hence, in LVK analyses, only $\delta \kappa_s$ is measured along with the other binary parameters, while assuming $\delta \kappa_a = 0$ (which is true for BBHs). This is a strong assumption, implying that this approach will only be applicable to binaries consisting of compact objects with identical spin-induced deformations. Nevertheless, a follow-up investigation will be needed if the data suggest a $\delta \kappa_s$ posterior significantly different from zero. This method of constraining $\delta \kappa_s$ (and $\delta \kappa_a$) using Bayesian inference was first developed in Krishnendu et al.~\cite{Krishnendu:2019tjp}, where only the $2$PN and $3$PN phase corrections due to the spin-induced quadrupole moment were included. 

This test was first applied to the GWTC-2 events
that have an SNR $\geq6$ in the inspiral phase estimated using a GR BBH waveform: 17 events satisfied this condition. Using the IMRPhenomPv2 for all events and IMRPhenomPv3HM for GW190412, the analysis found a combined bound on $\delta \kappa_s$ to be  $-23.2^{+52.2}_{-62.4}$. For GWTC-3 analysis, the selection criteria were tightened a bit --- along with the SNR threshold the events also required their effective inspiral spin parameter $\chi_{\rm eff}$  posterior from standard parameter estimation to exclude zero at the 68\% credible interval. This is because this test relies on the presence of spin in at least one compact object in the binary, otherwise there would not be any spin-induced moments. Thus, the criterion on $\chi_{\rm eff}$ makes sure that the test is applied to systems with significant spin to obtain meaningful constraints on $\delta \kappa_s$. Only 13 events in GWTC-3 passed these two criteria. Using IMRPhenomPv2 for all selected events, the combined bound on $\delta \kappa_s$ was found to be $-26.3^{+45.8}_{-52.9}$, consistent with the Kerr BBH population hypothesis.

\begin{figure*}[htb]
\centering
\includegraphics[width=\textwidth]{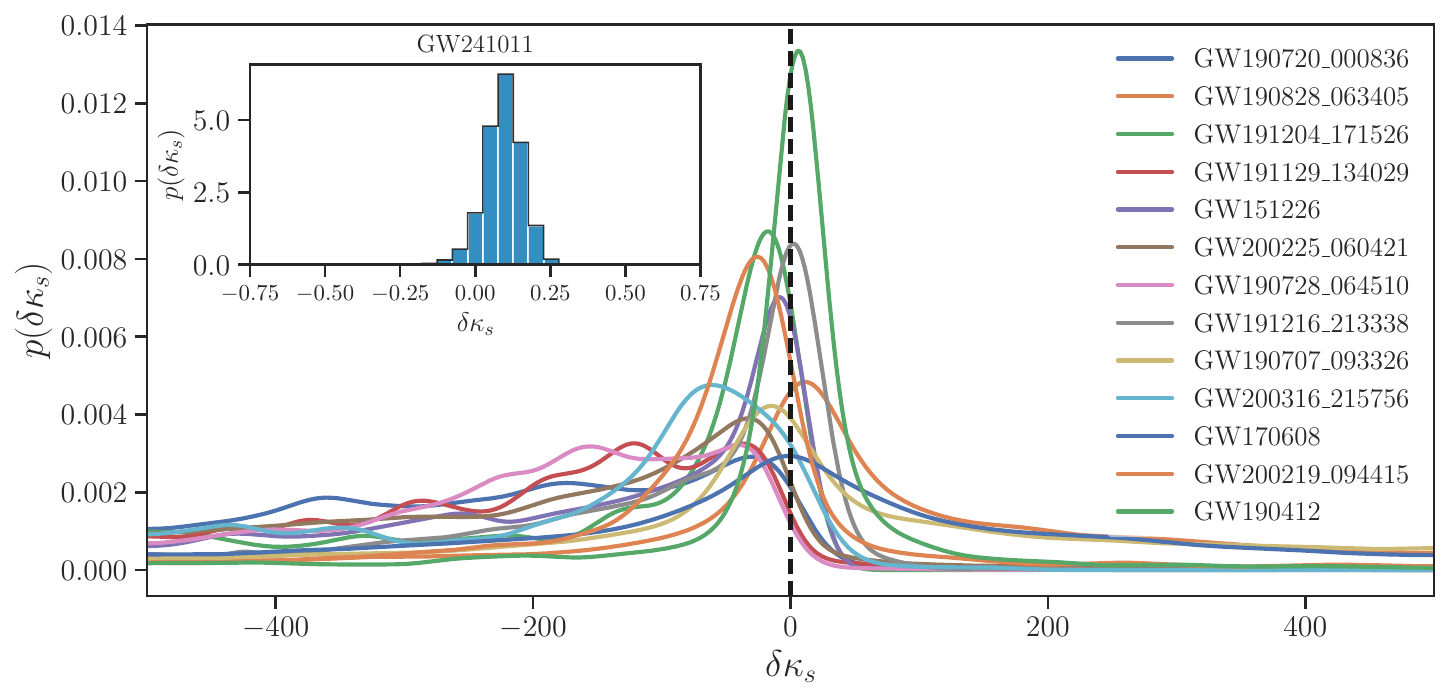}
\caption{\label{fig:siqm}Marginalized posterior on $\delta \kappa_s$ for 13 GWTC-3 events analyzed using IMRPhenomPv2. The inset shows the $\delta \kappa_s$ posterior for GW241011, which was analyzed using IMRPhenomXPHM. Note the x-axis range and the markedly tighter constraints on $\delta \kappa_s$ for GW241011, compared to the GWTC-3 events. This figure is prepared using public data and codes from~\cite{LIGOScientific:2021sio,LIGOScientific:2025brd,ligo_scientific_collaboration_virgo_coll_2022_7007370,ligo_scientific_collaboration_2025_17343574}.}
\end{figure*}

Later, Divyajyoti et al.~\cite{Divyajyoti:2023izl} extended the framework to use IMRPhenomXPHM, which incorporates two-spin precession and higher harmonics, as the baseline GR waveform model, and reanalyzed several GWTC-3 events that previously provided better bounds on $\delta \kappa_s$ with IMRPhenomPv2. Specifically, they analyzed GW151226, GW170608, GW190412, and GW191204\_171526 using IMRPhenomXPHM and found significantly tighter constraints for unequal-mass events such as GW190412. However, all four events analyzed in~\cite{Divyajyoti:2023izl} were consistent with the Kerr BBH hypothesis. The recent event GW241011~\cite{LIGOScientific:2025brd}, characterized by a high primary spin and mass asymmetry, has provided the most stringent constraints to date with $\delta \kappa_s = 0.10^{+0.09}_{-0.11}$ obtained using the IMRPhenomXPHM waveform model. This measurement surpasses the previous best bound from GW190412, which reported $\delta \kappa_s = 0^{+2}_{-91}$~\cite{Divyajyoti:2023izl}. In figure~\ref{fig:siqm}, we show the posterior distributions of $\delta \kappa_s$ for 13 GWTC-3 events and GW241011. The posterior for GW241011 is significantly narrower than those for the GWTC-3 events.

Saini et al.~\cite{Saini:2023gaw} explored the prospect of constraining deviations in spin-induced octupole moments from the Kerr nature while keeping the spin-induced quadrupole moments fixed to their Kerr values (i.e., $\delta \kappa_s = \delta \kappa_a = 0$). Using the IMRPhenomPv2 model, they analyzed seven events from GWTC-3 but found no meaningful constraints on the spin-induced octupole moments. They attributed these weak constraints to the small spins of the events and the limited sensitivity of the LIGO/Virgo detectors.

In the spirit of inspiral tests of Kerr nature, several authors have explored beyond-Kerr spacetimes and placed bounds on deviations from the Kerr metric using GW observations. For example, Carson et al.~\cite{Carson:2020iik} derived the modifications to the GW phase due to the Johannsen-Psaltis~\cite{Johannsen:2011dh} metric, which appear at $2$PN order and can be parameterized by a single deviation parameter. Santos et al.~\cite{Santos:2024pfa} used these modifications to constrain the deviation parameter of the Johannsen-Psaltis metric using 15 events from GWTC-3, employing IMRPhenomXPHM as their baseline GR model. They found the deviation parameter to be consistent with zero for all 15 events, supporting the Kerr spacetime hypothesis. Das et al.~\cite{Das:2024mjq} derived the phase corrections due to the Johannsen~\cite{Johannsen:2013szh} metric up to the $4$PN order, which are parameterized in terms of two deviation parameters. They analyzed six GWTC-3 events with total mass $<80M_\odot$ and network SNR $>15$, again using IMRPhenomXPHM as the baseline model. They found the analyzed events to be consistent with Kerr spacetime when only one deformation parameter was constrained (while keeping the other parameter fixed to zero). However, no meaningful bounds could be placed when both deformation parameters were allowed to vary simultaneously.

\subsubsection{Ringdown tests}
Binary coalescences involving black holes or neutron stars initially produce a highly deformed black hole\footnote{Depending on the mass of the remnant, mergers of binary neutron stars could also result in an infinitely stable neutron star~\cite{Sarin:2020gxb}.} which soon after settles down to a quiet state of ringdown, emitting gravitational radiation in the form of quasi-normal modes (QNMs)~\cite{Vishveshwara:1970zz,Press:1971wr} of complex frequencies. The real part of the complex 
frequency represents the oscillation frequency while the imaginary part represents the inverse of the damping time of the 
mode. The ringdown spectrum is typically modeled as a combination of exponentially damped sinusoids~\cite{LIGOScientific:2020tif,LIGOScientific:2021sio}
\begin{equation}
\begin{aligned}
	h_{+}(t) - i h_{\times}(t) = \sum_{\ell = 2}^{+\infty} \sum_{m = - \ell}^{\ell} \sum_{n = 0}^{+\infty} \; & \; \mathcal{A}_{\ell m n} \; \exp \left[ -\frac{t-t_0}{(1+z)\tau_{\ell m n}} \right] \\
    & \times \exp \left[ -\frac{2\pi i f_{\ell m n}(t-t_0)}{1+z} \right]  {}_{-2}S_{\ell m n}(\theta, \phi, \chi_{\rm f}),
\end{aligned}
\end{equation}
where the indices $(\ell, m)$ label the angular multipoles, and $n$ denotes the overtone number, ordering modes of a given $(\ell, m)$ by decreasing damping time. The amplitude, frequency, and damping time of each mode are denoted by $\mathcal{A}_{\ell m n}$, $f_{\ell m n}$, and $\tau_{\ell m n}$, respectively. Here, $\chi_{\rm f}$ is the dimensionless spin of the remnant, $z$ is the source redshift, and $t_0$ is the start time of the ringdown model. Finally, ${}_{-2}S_{\ell m n}$ denotes the spin-weight $-2$ spheroidal harmonic basis functions, with $\theta$ and $\phi$ specifying the polar and azimuthal angles of the final spin axis relative to the observer.

The fundamental mode, given by $\ell=2, m=2, n=0$, is the least damped QNM~\cite{Detweiler:1980gk,Dreyer:2003bv,Berti:2005ys} and dominates the ringdown radiation at the late-time. GR predicts a unique relationship between the mass and spin of a Kerr black hole and the QNM frequencies and damping times~\cite{Vishveshwara1970b,1971ApJ...170L.105P,Teukolsky:1973ha,Chandrasekhar:1975zza}. Thus, measuring the QNM frequencies and damping times allows testing the {\it no-hair conjecture}~\cite{Hansen:1974zz,Carter:1971zc,Gurlebeck:2015xpa} which states that a Kerr black hole can be completely characterized by its mass and spin\footnote{assuming electric charge is negligible for astrophysical black holes~\cite{Blandford:1977ds,Hanni:1982bc,Gibbons:1975kk}.}. Hence, the detection of {\it multiple} QNMs and checking the consistency between the inferred mass and spin of the compact object is the premise of {\it black hole spectroscopy} that not only verifies the no-hair conjecture but also tests the nature of the binary merger remnant. 

The detection of GW150914 made it possible to perform the ringdown test for the first time. A simple damped-sinusoid signal model was used to test the consistency of the data with the fundamental QNM of the remnant of GW150914~\cite{LIGOScientific:2016lio}. By varying the QNM start time, it was concluded that the GW150914 data were consistent with the presence of the fundamental QNM as predicted by GR. Later, GW150914 became a testbed for ringdown analyses by different independent groups~\cite{Carullo:2019flw,Isi:2019aib,Isi:2021iql,Cotesta:2022pci,Bhagwat:2017tkm,Baibhav:2023clw,Isi:2022mhy,Cotesta:2022pci,Carullo:2023gtf,Baibhav:2023clw,Isi:2023nif,Finch:2022ynt,Ma:2023cwe,Ma:2023vvr,CalderonBustillo:2020rmh,Wang:2023ljx,Correia:2023bfn}, who examined the full ringdown spectrum using various data analysis techniques (such as time-domain versus frequency-domain approaches) and data handling choices (such as the selection of the ringdown start time), but they could not agree on the presence of more than one overtone in GW150914 data. We refer the reader to section 6.2.2 of~\cite{Berti:2025hly}, which summarizes the debate between these groups, and concludes that one needs a larger ringdown SNR than in GW150914 to confirm the presence of multiple QNMs confidently.

Within the LVK Collaboration, ringdown tests using the full spectrum were performed for events in GWTC-2~\cite{LIGOScientific:2020tif} and GWTC-3~\cite{LIGOScientific:2021sio} through two complementary approaches: {\it pyRing}~\cite{Carullo:2019flw,Isi:2019aib} and {\it pSEOBNRv4HM}~\cite{Ghosh:2021mrv}. The pyRing method employs a time-domain damped-sinusoid waveform model (along with a time-domain likelihood function) to analyze only the ringdown portion of the signal. The ringdown start time is set to the peak of $(h_+^2 + h_\times^2)$, as determined from the full IMR parameter estimation assuming GR. In contrast, pSEOBNRv4HM uses a complete time-domain IMR waveform, combined with a frequency-domain likelihood function, to analyze the entire GW signal. In this model, the ringdown start time is inherently defined through calibrations with numerical relativity simulations. The pyRing method constrains fractional deviations from GR in the frequency ($\delta \hat{f}_{221}$) and damping time ($\delta \hat{\tau}_{221}$) of the $\ell=2, m=2, n=1$ mode, while assuming that the frequency and damping time of the $\ell=2, m=2, n=0$ mode remain consistent with GR predictions. Using 17 events from GWTC-2, the combined constraint on $\delta \hat{f}_{221}$ was $0.02^{+0.29}_{-0.33}$, consistent with all merger remnants being Kerr black holes. However, the constraints on $\delta \hat{\tau}_{221}$ were uninformative within the chosen prior bounds. With the increased number of events in GWTC-3 (21), the combined bound on $\delta \hat{f}_{221}$ improved slightly to $0.01^{+0.27}_{-0.28}$, again consistent with the Kerr black hole hypothesis. In this analysis, GW191109\_010717 was excluded because the final mass and spin estimates from the ringdown analysis showed a discrepancy with those obtained from the full IMR analysis. The constraints on $\delta \hat{\tau}_{221}$ again turned out to be uninformative, even with additional events in GWTC-3. The pSEOBNRv4HM method constrains fractional deviations in the frequency ($\delta \hat{f}_{220}$) and damping time ($\delta \hat{\tau}_{220}$) of the fundamental $\ell=2, m=2, n=0$ mode, while keeping all other QNMs fixed to their GR-predicted values. Combining results from nine events in GWTC-2 yielded $\delta \hat{f}_{220} = 0.03^{+0.38}_{-0.35}$ and $\delta \hat{\tau}_{220} = 0.16^{+0.98}_{-0.98}$, both consistent with GR predictions. The GWTC-3 analysis, which included twelve events, significantly improved these bounds to $\delta \hat{f}_{220} = 0.02^{+0.07}_{-0.07}$ and $\delta \hat{\tau}_{220} = 0.13^{+0.21}_{-0.22}$.

GWTC-2 included the event GW190521~\cite{LIGOScientific:2020iuh}, which attracted considerable attention as the most massive BBH merger detected at the time. The LIGO-Virgo Collaboration analyzed~\cite{LIGOScientific:2020ufj} this signal using pyRing and found no strong evidence for the presence of higher multipoles or overtones~\cite{LIGOScientific:2020ufj}. In contrast, Capano et al.~\cite{Capano:2021etf,Capano:2022zqm} reported evidence for the $\ell = 3, m = 3$ mode in the GW190521 data. Later, Siegel et al.~\cite{Siegel:2023lxl} reanalyzed the same data and were able to largely reproduce both the LIGO-Virgo Collaboration's~\cite{LIGOScientific:2020ufj} and Capano et al.'s~\cite{Capano:2021etf,Capano:2022zqm} findings. They also found support for the $\ell = 2, m = 1$ and $\ell = 2, m = 2$ modes and argued that the excitation of $\ell \neq m$ modes could be associated with the possible presence of precession~\cite{Miller:2023ncs,Zhu:2023fnf} or orbital eccentricity~\cite{Romero-Shaw:2020thy,Gayathri:2020coq} in GW190521.

Later, GW231123~\cite{LIGOScientific:2025rsn} emerged as the most massive BBH merger detected so far, which also exhibited high spin. The PyRing analysis of GW231123 found the merger remnant is consistent with a Kerr black hole~\cite{LIGOScientific:2025rsn}. However, Wang et al.~\cite{Wang:2025rvn} reported strong evidence for the presence of the $(\ell, m, n)=(2,2,0) + (2,0,0)$ modes in the GW231123 signal. Siegel et al.~\cite{Siegel:2025xgb} also analyzed the GW231123 signal and reported evidence for multiple QNMs, identifying $(\ell, m, n) = (2,1,0)$ as the relevant mode, in contrast to the $(2,0,0)$ mode found by Wang et al.~\cite{Wang:2025rvn}. Furthermore, due to its high SNR, GW250114 is the only event so far that confidently exhibits the presence of at least two QNMs. The signal is consistent with the first overtone being present at multiple times after the peak amplitude, and, for the first time, shows indications of the hexadecapolar ($\ell = 4$) mode frequency~\cite{LIGOScientific:2025obp}. On the other hand, Yang et al.~\cite{Yang:2025ror} reported statistically significant evidence for nonlinear quadratic $(\ell,m,n)=(2,2,0)\times(2,2,0)$ QNMs relative to the linear (4,4,0) mode in GW250114.

In the spirit of ringdown tests of the Kerr nature, Ahmed et al.~\cite{Ahmed:2024ykc} employed the Johannsen-Psaltis metric to investigate deviations from the Kerr spacetime using only the ringdown phase of the signal. They analyzed the ringdowns of GW150914 and GW190521 and found both events to be consistent with the Kerr metric.

\subsubsection{Echoes searches}
Certain alternative theories of gravity~\cite{Cardoso:2019rvt,Maggio:2021ans,PhysRev.172.1331,PhysRev.187.1767,Mathur:2009hf,Bena:2022rna,Mazur:2004fk,Bowers:1974tgi,Raposo:2018rjn,Alho:2022bki,1988AmJPh..56..395M,Visser:1995cc,Lemos:2003jb,Haensel:1986qb,Alcock:1986hz,Kouvaris:2015rea} predict the existence of compact objects that lack an event horizon but instead possess reflective surfaces. If the merger remnant is such a horizonless compact object, the ingoing GWs generated during the merger can be reflected multiple times between the effective potential barrier and the reflective surface. This process produces a series of repeated pulses that leak out to infinity, known as {\it echoes}. Thus, the detection of echoes in GW data would provide clear evidence for the existence of such horizonless exotic compact objects~\cite{Cardoso:2016rao,Cardoso:2016oxy,Maselli:2017tfq}.

The first search for GW echoes was carried out by Abedi et al.~\cite{Abedi:2016hgu}, who analyzed data from the first three GW events: GW150914, GW151226, and LVT151012 (now termed as GW151012). They searched for repeating damped echoes with time delays corresponding to Planck-scale deviations from GR near the event horizons and reported tentative evidence for their presence at a $2.5\sigma$ confidence level. However, this claim was later contested by several independent groups~\cite{Ashton:2016xff, Nielsen:2018lkf, Lo:2018sep, Uchikata:2019frs, Wang:2020ayy, Westerweck:2021nue}, who reanalyzed the data and found no significant evidence of echoes. Subsequently, Abedi et al.~\cite{Abedi:2018npz} analyzed the GW170817 data and claimed tentative evidence of echoes occurring about $1.0$ second after the merger, with a $4.2\sigma$ significance. In a later publication~\cite{Abedi:2020sgg}, the authors reviewed the status of echoes searches in GW data and argued that results from different groups could be reconciled if the echoes predominantly contribute at lower frequencies and/or originate from binary mergers with more unequal mass components.

Within the LVK Collaboration, searches for echoes were conducted starting from GWTC-2. For the GWTC-2 analysis, a morphology-dependent approach~\cite{Lo:2018sep} was used to examine 31 events. In this method, the ringdown portion of an IMR waveform model is modified to incorporate echoes, following the template proposed by Abedi et al.~\cite{Abedi:2016hgu}. Using the IMRPhenomPv2 waveform model (and NRSur7dq4 for GW190521) to generate echo signals,~\cite{LIGOScientific:2020tif} computed Bayes factors comparing hypotheses with and without echoes but found no statistically significant evidence for echoes in the GWTC-2 data. For the GWTC-3 analysis, a morphology-independent method~\cite{Tsang:2018uie, Tsang:2019zra} was employed using the BayesWave~\cite{Cornish:2014kda,Littenberg:2014oda}, which models potential echoes using sine-Gaussian basis functions. Again, no statistically significant evidence for echoes was found in the GWTC-3 data.

\section{Results from binaries containing neutron stars}
\label{sec:bns}
So far, our discussion has predominantly focused on BBH mergers. In this section, we present notable results and corresponding tests of GR from GW events involving neutron stars --- GW170817~\cite{LIGOScientific:2017vwq}, GW190425~\cite{LIGOScientific:2020aai}, GW200105\_162426, GW200115\_042309~\cite{LIGOScientific:2021qlt}, and GW230529\_181500~\cite{LIGOScientific:2024elc}.

{\bf GW170817} remains the only GW event to date observed in coincidence with an electromagnetic counterpart~\cite{LIGOScientific:2017ync}. This joint detection of GWs and gamma rays has provided remarkably stringent constraints on several aspects of fundamental physics~\cite{LIGOScientific:2017zic}. For instance, the observed time delay between the GW and gamma-ray signals constrained the difference between the speed of gravity and the speed of light to lie between $-3\times10^{-15}\,c$ and $+7\times10^{-17}\,c$. Furthermore, it improved existing bounds on local Lorentz invariance violation parameters by at least an order of magnitude relative to previous limits~\cite{Kostelecky:2015dpa,Shao:2014oha,Shao:2014bfa}. The observation also allowed a test of the Shapiro delay~\cite{Shapiro:1964pl} between gravitational and electromagnetic radiation, constraining deviations from Einstein-Maxwell theory to $-2.6\times10^{-7} \leq \gamma_{\rm GW}-\gamma_{\rm EM} \leq 1.2\times10^{-6}$, where $\gamma_{\rm GW}=1=\gamma_{\rm EM}$ in Einstein-Maxwell theory.

Subsequently, the LIGO-Virgo Collaboration performed comprehensive tests of GR on GW170817 while explicitly incorporating neutron star tidal effects into the waveform models~\cite{LIGOScientific:2018dkp}. Employing IMRPhenomPNRT~\cite{Dietrich:2017aum,Hannam:2013oca,Schmidt:2012rh,Schmidt:2014iyl} for the TIGER and SEOBNRT~\cite{Hinderer:2016eia,Steinhoff:2016rfi,Bohe:2016gbl,Barausse:2009xi,Barausse:2011ys,Taracchini:2013rva} for FTI, GW170817 yielded the first stringent constraints on the $-1$PN deviation parameter, which remains the most precise bound obtained from any GW observation to date. However, this limit is still weaker than that inferred from the binary pulsar system PSR J0737-3039~\cite{Lyne:2004cj,Kramer:2021jcw}. Conversely, the bound on the graviton mass derived from GW170817 was weaker than those obtained from BBH mergers --- an expected outcome, since GW170817 is the nearest source detected to date. For a fixed SNR, propagation-based tests tend to be more sensitive for more distant sources. The multimessenger observation of GW170817 also allowed independent measurements of the luminosity distance from both GW and electromagnetic data. Comparing these measurements placed constraints on the possibility of extra spacetime dimensions, which were found to be consistent with GR's four-dimensional prediction (see also~\cite{Pardo:2018ipy}). Owing to its precisely determined sky location, GW170817 provided the strongest evidence to date favoring purely tensor modes over purely scalar or purely vector modes. Finally, the multimessenger observation of GW170817 heavily constrained a range of beyond-GR cosmological models~\cite{Baker:2017hug} and dark energy models in beyond-GR theories~\cite{Creminelli:2017sry,Sakstein:2017xjx,Ezquiaga:2017ekz}.

{\bf GW190425} is the second candidate binary neutron star merger after GW170817. Since it was detected by only a single detector (LIGO Livingston), the event was not included in any of the GR tests conducted by the LVK Collaboration. Consequently, no significant results from tests of GR have been reported in the literature for this event.

{\bf GW200105\_162426 and GW200115\_042309} are two high-significance events detected during the third observing run, both consistent with being neutron star-black hole mergers. The initial parameterized test~\cite{LIGOScientific:2021qlt} found that neither event provided tighter constraints than those reported in the GWTC-2 analyses~\cite{LIGOScientific:2020tif}, although both were consistent with GR predictions. Subsequently, only GW200115\_042309 met the FAR threshold for inclusion in the GWTC-3 tests of GR analyses. Residual test, FTI, and echoes searches were performed for this event, assuming negligible matter effects, and all results were consistent with GR. Owing to its relatively long signal duration, GW200115\_042309 helped improve the constraint on the $-1$PN deviation parameter by a factor of two in the GWTC-3 analysis.

{\bf GW230529\_181500} is most likely a neutron star-black hole merger, with the primary component's mass falling within the so-called {\it low-mass gap}~\cite{Rhoades:1974fn,Kalogera:1996ci}, between typical neutron star and black hole masses ($\sim3$-$5M_\odot$). Since this event was detected by only a single interferometer (LIGO Livingston), it will not be included in the upcoming GWTC-4 tests of GR analyses. Nevertheless, TIGER and FTI analyses were carried out in the discovery paper~\cite{LIGOScientific:2024elc} and by S{\"a}nger et al.~\cite{Sanger:2024axs}. Owing to its relatively low total mass, this event yielded an exceptionally tight constraint on the $-1$PN deviation parameter --- about $17$ times more stringent than that obtained from the earlier neutron star-black hole event GW200115\_042309. When mapped to a class of Einstein-scalar-Gauss-Bonnet gravity theories, the $-1$PN result provided the most stringent upper bound to date on the Gauss-Bonnet coupling constant, $\sqrt{\alpha_{\rm GB}} \lesssim 0.28\,{\rm km}$ (see~\cite{Wang:2025ehy} for constraints on Screened Modified Gravity and Brans-Dicke theory derived from GW230529\_181500).

\section{Summary and Future Prospects}
\label{sec:concl}
The first decade of GW observations has delivered a wealth of groundbreaking detections, offering unprecedented opportunities to test fundamental physics under extreme conditions, which were not possible before. Several exceptional events have allowed remarkably precise constraints on the predictions of GR and several beyond-GR theories. GW150914 marked the beginning of GW-based tests of GR, allowing the first constraints on deviations from post-Newtonian theory and on the graviton's mass. GW170814 made it possible, for the first time, to probe the polarization content of GWs using observations from three detectors. The multimessenger detection of GW170817 opened new avenues for constraining the difference between the speeds of gravity and light, constraining the Shapiro delay to test the equivalence principle, tightening bounds on local Lorentz invariance violations, and ruling out several beyond-GR cosmological models. GW241011 provided the most stringent constraints yet on the spin-induced quadrupole moment and on deviations in the amplitude of the $(\ell, m) = (3, \pm3)$ spherical harmonic modes predicted by GR~\cite{Puecher:2022sfm,LIGOScientific:2025brd,Gupta:2025paz}, surpassing previous benchmarks set by GW190814 and GW190412~\cite{Puecher:2022sfm}. Finally, GW250114 marked the first instance where an overtone could be confidently detected, facilitating tests of the Kerr nature of black holes through the identification of multiple QNMs in the ringdown spectrum of the merger remnant, and showing remarkable consistency with Hawking's area law~\cite{Hawking:1971tu}.

Despite this tremendous progress, we are still only at the beginning of rigorous precision tests of GR with GWs. Detecting a confident and real deviation from GR remains one of the biggest challenges in modern physics. As highlighted in~\cite{Gupta:2024gun}, several improvements are still needed in both waveform modeling and data analysis techniques --- otherwise, apparent violations of GR may arise from modeling or instrumental limitations rather than genuine physics. For instance, neglecting effects such as orbital eccentricity~\cite{Narayan:2023vhm,Saini:2022igm,Bhat:2022amc,Shaikh:2024wyn}, strong lensing~\cite{Narayan:2024rat,Wright:2024mco}, microlensing~\cite{Mishra:2023vzo}, millilensing~\cite{Liu:2024xxn}, or higher harmonics~\cite{Pang:2018hjb} in waveform models can mimic apparent deviations from GR. Similarly, unaccounted non-Gaussian~\cite{Kwok:2021zny} or non-stationary noise can distort our inferences, making the data appear inconsistent with GR for the wrong reasons. Hence, to move forward, it is essential to incorporate missing physical effects into waveform models wherever possible, quantify potential biases when that is not feasible, and develop a robust framework to separate false GR violations from genuine ones. The next few years promise to be particularly exciting: with improving detector sensitivity, more sophisticated models, and richer data, we may finally begin to determine whether GR remains a valid description of gravity in the strong-field, highly dynamical regime, or whether deviations from Einstein's theory emerge under such extreme conditions.

\section*{Acknowledgments}
We thank K.~G.~Arun for carefully reading the paper and providing useful comments. We also thank the LIGO-Virgo-KAGRA Collaboration's Testing GR group members for implementing and reviewing the GR tests, and for their contributions to the analyses and to the publication of the results that we discussed in this paper. This document has LIGO preprint number P2500706.

AG is supported by NSF grants PHY-2308887 and CAREER-2440327. AG is grateful for the hospitality of Perimeter Institute where this paper was written. Research at Perimeter Institute is supported in part by the Government of Canada through the Department of
Innovation, Science and Economic Development and by the Province of Ontario through the Ministry of Colleges and Universities. This work was supported by a grant from the Simons Foundation (1034867, Dittrich). 

LIGO Laboratory and Advanced LIGO are funded by the United States National Science Foundation (NSF) as well as the Science and Technology Facilities Council (STFC) of the United Kingdom, the Max-Planck-Society (MPS), and the State of Niedersachsen/Germany for support of the construction of Advanced LIGO and construction and operation of the GEO600 detector. Additional support for Advanced LIGO was provided by the Australian Research Council. Virgo is funded, through the European Gravitational Observatory (EGO), by the French Centre National de Recherche Scientifique (CNRS), the Italian Istituto Nazionale di Fisica Nucleare (INFN) and the Dutch Nikhef, with contributions by institutions from Belgium, Germany, Greece, Hungary, Ireland, Japan, Monaco, Poland, Portugal, Spain. The construction and operation of KAGRA are funded by Ministry of Education, Culture, Sports, Science and Technology (MEXT), and Japan Society for the Promotion of Science (JSPS), National Research Foundation (NRF) and Ministry of Science and ICT (MSIT) in Korea, Academia Sinica (AS) and the Ministry of Science and Technology (MoST) in Taiwan. Unless otherwise specified, the contents of this release are licensed under the Creative Commons Attribution 4.0 International License. To view a copy of this license, visit http://creativecommons.org/licenses/by/4.0/ or send a letter to Creative Commons, PO Box 1866, Mountain View, CA 94042, USA.


\bigskip

\bibliographystyle{iopart-num}
\bibliography{gr_review}

\end{document}